\documentclass[aps,prb,preprint,groupedaddress,showpacs]{revtex4-1}

\usepackage{graphicx}
\input{epsf}

\begin{document}
\title{Electronic
inhomogeneities in the superconducting phase of
CaFe$_{1.96}$Ni$_{0.04}$As$_{2}$ single crystals}

\author{Anirban Dutta$^1$, Neeraj Kumar$^2$, A. Thamizhavel$^2$, and Anjan K. Gupta$^1$}
\address{$^1$Department of Physics, Indian Institute of
Technology Kanpur, Kanpur 208016, India\\ $^2$Department of
Condensed Matter Physics and Materials Science, Tata Institute of
Fundamental Research, Homi Bhabha Road, Colaba, Mumbai 400 005,
India}

\begin{abstract}
Superconductivity in CaFe$_{2-x}$Ni$_{x}$As$_{2}$ emerges in close
proximity to an antiferromagnetic (AFM) ordered parent state and the
AFM phase overlaps with superconducting (SC) phase for a small range of \emph{x}-values. We present scanning tunneling microscopy and spectroscopy study of an underdoped CaFe$_{2-x}$Ni$_{x}$As$_{2}$ single crystal in the vicinity of the boundary of the two phases. Both resistivity and magnetic susceptibility measurements show a superconducting T$_C$ of 15 K and from later we deduce a superconducting fraction of 1.2 $\%$. Topographic images show reasonably flat surface with signatures of atomic resolution. Spectra between 120 K and 20 K  are spatially homogeneous and show signatures of
spin density wave (SDW) gap. Below $T_{C}$, spectra show significant spatial inhomogeneity with a depression in density of states in $\pm$ 5 meV energy range. Inhomogeneity reduces significantly as the temperature goes above $T_{C}$ and disappears completely far above $T_C$. These observations are discussed in terms of an inhomogeneous electronic phase that may exist due to the vicinity of this composition to the SC dome boundary on the underdoped side of the phase diagram.
\end{abstract}


\maketitle
\section{Introduction}
The superconductivity in Iron-Arsenic (Fe-As) based
pnictides \cite{kamihara} emerges in close proximity to an AFM ordered parent state and $T_C$ has dome-shaped dependence on doping or pressure
\cite{luetkens,zhao,peng,goko,park,erwin,chen,marsik,bernhard,drew,sana,laplace}.
In some pnictides, the AFM and SC phases overlap in the phase
diagram and then the maximum $T_C$ is found close to the extrapolated end
point of the AFM transition. It is widely believed that magnetic fluctuations have an important role in the origin of high-$T_C$ superconductivity. Quantum fluctuations
associated with the quantum critical point
(QCP)\cite{zhou,shibauchi} may also have a crucial role in
superconductivity. Muon spin relaxation ($\mu$SR) and
M$\ddot{o}$ssbauer spectroscopy on LaFeAsO$_{1-x}$F$_x$ show a
discontinuous first-order-like transition from SDW to SC state
without any coexistence of SDW and SC phases \cite{luetkens}.
Neutron scattering in CeFeAsO$_{1-x}$F$_x$ reveals a continuous
second order transition, but the SDW and SC phases touch only at T =
0 and this could be a quantum critical point \cite{zhao}. On the
contrary, in other cases such as scanning tunneling microscopy and
spectroscopy on NaFe$_{1-x}$Co$_x$As \cite{peng}, $\mu$SR
\cite{goko,park,erwin} and neutron diffraction \cite{chen} study on
Ba$_{1-x}$K$_x$Fe$_2$As$_2$, $\mu$SR study on
Ba(Fe$_{1-x}$Co$_x$)$_2$As$_2$ \cite{marsik,bernhard} and on SmFeAsO$_{1-x}$F$_x$ \cite{drew}, $\mu$SR and nuclear quadropole
resonance study on SmFe$_{1-x}$Ru$_x$AsO$_{0.85}$F$_{0.15}$
\cite{sana} and $^{75}$As NMR study on
Ba(Fe$_{1-x}$Ru$_x$)$_2$As$_2$ \cite{laplace} show coexistence of
the two phases. These two phases may coexist microscopically or in a
phase separated way. SmFe$_{1-x}$Ru$_x$AsO$_{0.85}$F$_{0.15}$
\cite{sana} and Ba(Fe$_{1-x}$Ru$_x$)$_2$As$_2$ \cite{laplace}
display phase separation while NaFe$_{1-x}$Co$_x$As \cite{peng},
Ba(Fe$_{1-x}$Co$_x$)$_2$As$_2$ \cite{marsik,bernhard} show
microscopic coexistence. In some pnictides for example
Ba$_{1-x}$K$_x$Fe$_2$As$_2$ and SmFeAsO$_{1-x}$F$_x$, the issue of
whether the two phases, AFM and SC, coexist microscopically or as phase separated, is not completely clear \cite{goko,park,chen,erwin,drew}. With its atomic scale structural and spectroscopic imaging capabilities, scanning tunneling microscopy and spectroscopy (STM/S) is an ideal probe to investigate the local electronic properties of
these systems. In our variable temperature STM/S we investigate
the temperature evolution of the electronic density of states (DOS) to see how
they correlate with various phases at different temperatures.

The parent compound CaFe$_{2}$As$_{2}$ undergoes a spin density wave
(SDW) transition near T$_{SDW}$ = 170 K \cite{nandi,ronning}. Around
T$_{SDW}$ a structural transition is also observed, where the
symmetry changes from tetragonal ({\it I4/mmm}) to orthorhombic
({\it Fmmm})\cite{nandi,ronning}. Electron doping by partially
replacing Fe by Ni suppresses the SDW and the structural
transitions, leading to superconductivity in CaFe$_{2}$As$_{2}$
\cite{neeraj}. With increase of Ni concentration, T$_{SDW}$ starts to
decrease. Furthermore a drop in resistivity occurs at $\sim$ 15 K
for \emph{x} = 0.027 which at higher doping develops into a pure
superconducting transition \cite{neeraj}. SDW phase vanishes
completely at \emph{x} = 0.06 \cite{neeraj}.

In this paper, we report temperature dependent STM/S studies of underdoped CaFe$_{1.96}$Ni$_{0.04}$As$_{2}$ single crystals in 5.4 K - 292 K temperature range. We observed flat terraces with some signatures of atomic resolution. Spectra show homogeneous tunneling DOS with SDW gap at high temperature and below T$_{C}$ we see inhomogeneous local tunneling DOS with two kinds of spectra, either showing only SDW gap or a suppression in DOS in $\pm$ 5 mV bias range. Although these spectra below T$_C$ are unlike those of typical SC, but the low energy DOS suppression is observed only below T$_C$. A preliminary version of this work was also reported by us in a conference \cite{anirban1}.

\section{Experimental details}
Single crystals of CaFe$_{1.96}$Ni$_{0.04}$As$_{2}$ were grown
\cite{neeraj} by the high temperature solution growth using Sn-flux
under identical conditions as mentioned in Ref. \cite{neeraj2}.
Electrical resistivity was measured using a standard four-probe
method in a closed cycle refrigerator. DC magnetization was
measured in a superconducting quantum interference device
magnetometer. STM/S studies were done in cryogenic vacuum using a
homemade variable temperature STM with fresh-cut
Pt$_{0.8}$Ir$_{0.2}$ tips with RHK electronics and
software. For STM/S measurements crystals were cleaved \emph{in situ} at room temperature and at 4 $\times$ 10$^{-6}$ mbar pressure before transferring the
sample to the STM head at low temperature. Standard ac-modulation technique was used for STS measurements with a modulation amplitude between 1 and 10 mV and
frequency 2731 Hz. For small bias (20 mV) we plot dI/dV directly to show the DOS,
while we normalize the spectra for large bias (250 mV) to normalize away the effect of the tunnel matrix element on the spectra \cite{anirban}. This process also sharpens the spectral features. We have seen similar results on 4-5 different samples of the same composition with more than one tip.

\section{Results and Analysis}
\begin{figure}
\includegraphics[width=2.5in]{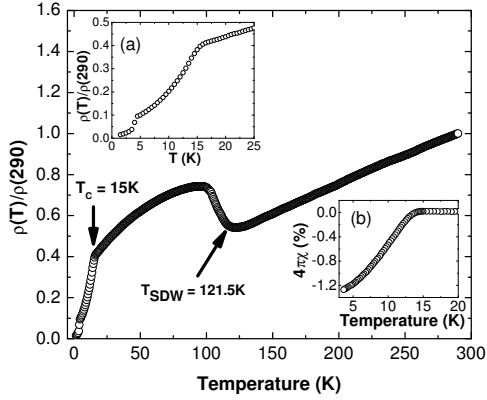}
\caption{\label{fig:RnChi} Temperature dependence of the in
plane electrical resistivity of a CaFe$_{1.96}$Ni$_{0.04}$As$_{2}$ single crystal. The inset (a) shows an
expended view of the region near the superconducting transition. The inset (b) shows temperature dependence of the zero field
cooling (ZFC) magnetic susceptibility at 50 Oe.}
\end{figure}

\mbox{Fig.} \ref{fig:RnChi} shows the in-plane electrical
resistivity of as-grown CaFe$_{1.96}$Ni$_{0.04}$As$_{2}$ single
crystal. The in-plane resistivity decreases slowly with decreasing
temperature below room temperature down to 121.5 K. Then there is a
relatively broad upturn starting at $T_{SDW} = 121.5 K$ due to SDW
transition. This rise in resistivity is attributed to the opening of
a gap in parts of the Fermi surface giving rise to a loss in DOS at
$E_F$ upon entering the SDW state. However, resistivity starts
decreasing rapidly as the temperature is reduced further indicating
that the Fermi surface is only partially gapped in the SDW state. As
the temperature decreases further, a sharp drop in the resistivity
occurs at $\sim$ 15 K due to the onset of superconducting order. But
the resistivity does not go to zero down to 1.3 K. Full
superconducting transition develops only at slightly higher Ni
doping (0.053 and 0.06)\cite{neeraj} with an onset $T_C$ of 15 K.
The drop in the resistivity at $\sim$ 4 K is due to the presence of
traces of Sn, which is used as a flux. The inset (b) of \mbox{Fig.} \ref{fig:RnChi} shows the temperature dependence of magnetic susceptibility $\chi$ measured under zero field cool (ZFC) condition at 50 Oe. Field was applied parallel to the ab-plane of
the crystals. The susceptibility remains unchanged (nearly zero)
with decreasing temperature down to 15 K. $\chi$ starts decreasing
for temperature below $\sim$ 15 K and becomes negative, but 4$\pi
\chi$ does not saturate down to 4 K. This indicates
that the system is not fully shielded. Assuming demagnetization
factor to be zero for $H \parallel ab$-plane, measured shielding
fraction is only $\sim$ 1.2 $\%$ at 5 K. Thus only a small portion
of the system is superconducting at 4 K.

\begin{figure}
\includegraphics[width=2.5in]{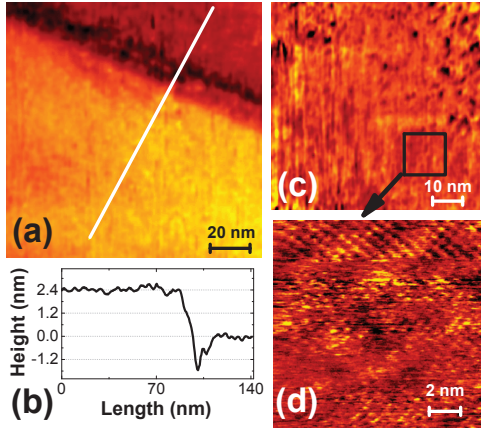}
\caption{\label{fig:image-292-5K}(Color online) (a) STM topographic image (area:
140$\times$140 nm$^2$) of CaFe$_{1.96}$Ni$_{0.04}$As$_{2}$ single
crystal at 292 K taken with a junction bias of 500 mV and a tunnel
current of 60 pA. (b) Topographic profile along the lines marked in (a). (c) and (d) are the STM topographic images at 5.4 K of 69.22 $\times$ 69.22 nm$^{2}$ and
13.17 $\times$ 13.17 nm$^{2}$ respectively. Image in (c) was taken
with a junction bias of 100 mV and a tunnel current of 100 pA
while that in (d) was taken with constant height mode. }
\end{figure}

\begin{figure}
\includegraphics[width=2.5in]{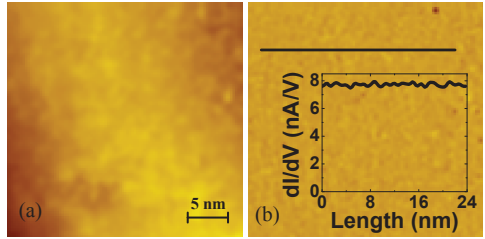}
\caption{\label{fig:toponcond-110K} (Color online) Simultaneously
acquired (a) topographic and (b) conductance images (area:
29.03$\times$29.03 nm$^2$) of CaFe$_{1.96}$Ni$_{0.04}$As$_{2}$ single
crystal at 110 K taken with a junction bias of 10 mV, a tunnel
current of 100 pA and an ac-modulation voltage of 1 mV. The inset in (b) shows the $dI/dV$ variation along the marked line, which is less than 2$\%$.}
\end{figure}

\mbox{Fig.} \ref{fig:image-292-5K} show STM topographic images of
\emph{in situ} cleaved CaFe$_{1.96}$Ni$_{0.04}$As$_{2}$ single
crystal at highest and lowest studied temperatures. Topographic image at 292 K (\mbox{Fig.} \ref{fig:image-292-5K} (a)) show atomically flat surface over the terraces, with an rms roughness over any terrace as $\sim$ 0.11 nm. Similar kind of topographic images are observed at all studied temperatures. The line profile in \mbox{Fig.} \ref{fig:image-292-5K} (b) shows that the terraces are separated by $\sim$ 2.4 ($\pm$0.1) nm, i.e. twice the atomic steps \cite{neeraj}.  Topographic images at the lowest studied temperature (5.4 K) are shown in \mbox{Fig.} \ref{fig:image-292-5K} (c) and (d), which also show signatures of atomic resolution. Simultaneously acquired STM topographic and conductance images (area: 29.03$\times$29.03 nm$^2$) at 110 K are shown in \mbox{Fig.} \ref{fig:toponcond-110K}. dI/dV along the marked line, in the
conductance map as plotted in the inset, shows very little variation indicating an electronically homogeneous surface.

\begin{figure}
\includegraphics[width=2.5in]{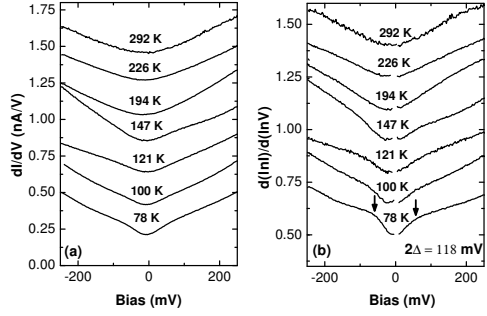}
\caption{\label{fig:spectra-SDW} Temperature dependent (a) $dI/dV$
and (b) $d(lnI)/d(lnV)$ spectra of CaFe$_{1.96}$Ni$_{0.04}$As$_{2}$
single crystals between 292 K and 78 K. Spectra were taken with a
junction bias of 250 mV, a tunnel current of 100 pA and an ac-modulation voltage of 10 mV. Consecutive spectra have been shifted uniformly upwards for clarity in all the three plots.}
\end{figure}

\begin{figure}[h!]
\includegraphics[width=2.5in]{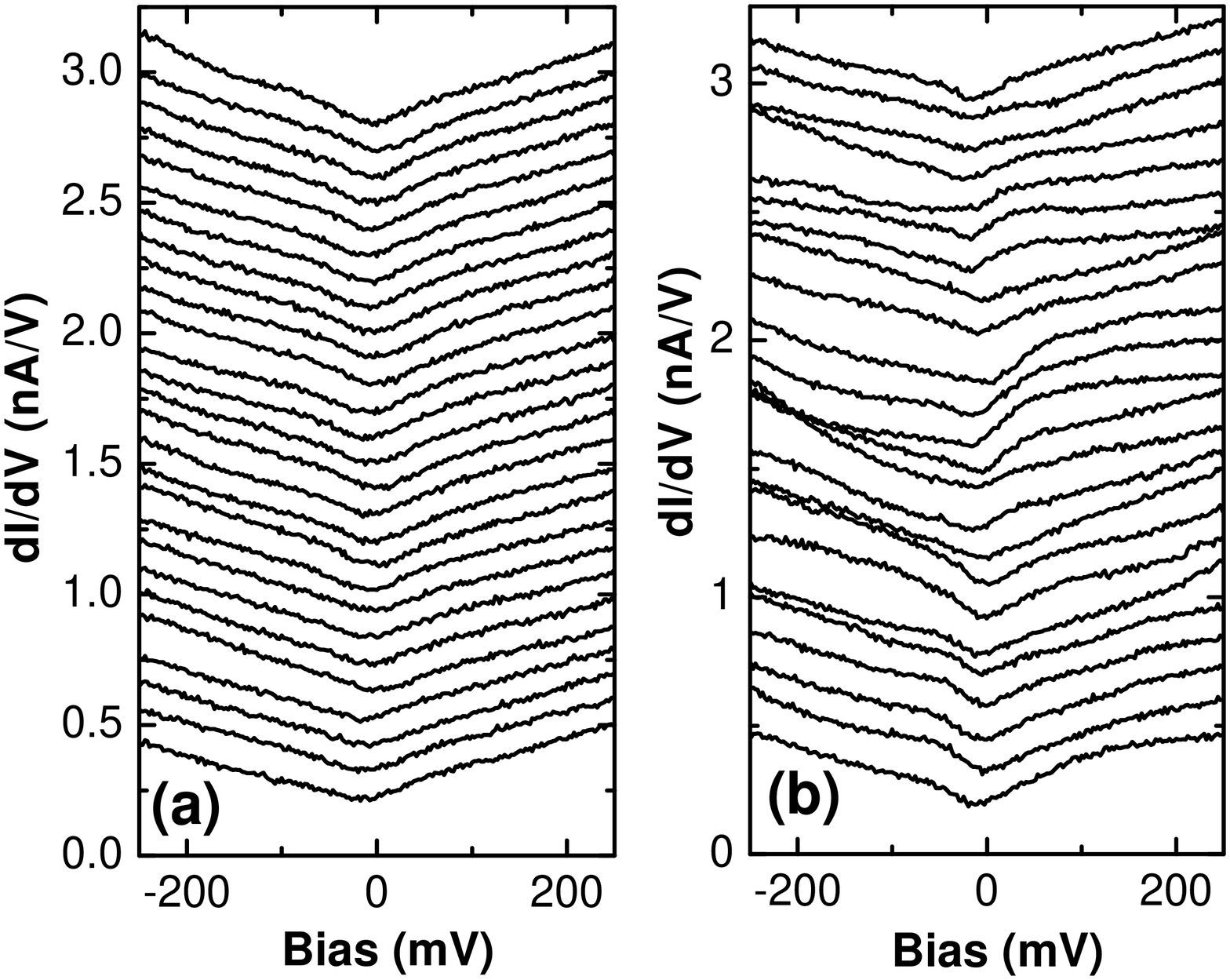}
\caption{\label{fig:spectra-78-5K} $dI/dV$ spectra of CaFe$_{1.96}$Ni$_{0.04}$As$_{2}$
single crystals taken at different locations of the sample at (a) 78 K and (b) 5.4 K. Spectra were taken with a junction bias of 250 mV, a tunnel current of 100 pA and an ac-modulation voltage of 10 mV. Consecutive spectra have been shifted uniformly upwards for clarity.}
\end{figure}

The temperature dependent tunneling spectra, dI/dV versus V between
78 K and 292 K are shown in \mbox{Fig.}
\ref{fig:spectra-SDW} (a). Each plotted spectrum is a spatial average of about one hundred spectra taken over an area of $\sim$ 2$\times$2 $\mu$m$^2$, as they show very little variations (See \mbox{Fig.}
\ref{fig:spectra-78-5K} (a)). d(lnI)/d(lnV) - V is plotted in \mbox{Fig.} \ref{fig:spectra-SDW} (b) which eliminates the effect of the voltage
dependence of the tunneling matrix element and sharpens the gap
feature \cite{anirban}. Above $T_{SDW}$ (121.5 K), there is a
broad depression in the d(ln I)/d(ln V)-V spectra which becomes more
pronounced as the temperature goes below $T_{SDW}$. We attribute
this to the opening of a partial gap at the $E_F$ upon entering the
SDW state. Similar kind of spectra were seen in EuFe$_2$As$_2$ \cite{anirban}. Local spectra at 78 K and 5.4 K, taken with same junction resistance and ac modulation are plotted in \mbox{Fig.} \ref{fig:spectra-78-5K}. We see significant spatial inhomogeneity at 5.4 K as compared to to that at 78 K.

\begin{figure*}
\includegraphics[width=5in]{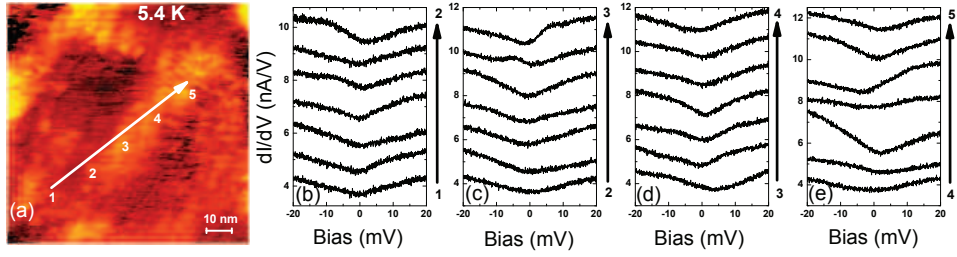}
\caption{\label{fig:spectra-along-line}(Color online) (a) STM
topographic image at 5.4 K. (b)-(e) Spatially resolved dI/dV spectra
along the marked line of the topographic image in (a). Separation
between two consecutive spectra is 2.4 nm. Spectra were taken with a
junction bias of 20 mV bias, a tunnel current of 100 pA and an ac-modulation voltage of 1 mV. Consecutive spectra have been shifted uniformly upwards for clarity in the plots (b)-(e).}
\end{figure*}

\begin{figure*} [h!]
\includegraphics[width=5in]{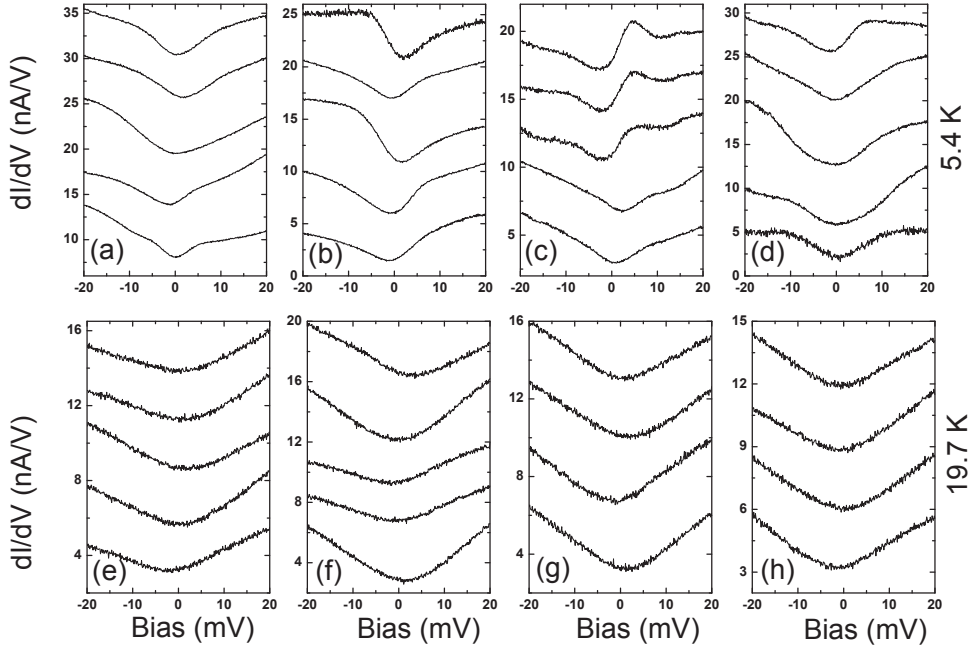}
\caption{\label{fig:spectra-5.4K-19.7K} (a)-(d) dI/dV spectra at
different locations taken with a junction bias of 20 mV, a tunnel
current of 100 pA and an ac-modulation voltage of 1 mV at 5.4 K. They have been grouped in different panels based on similarity between spectra. (e)-(h) dI/dV spectra at
different locations taken with same parameters but at 19.7
K. Consecutive spectra have been shifted uniformly upwards for
clarity in all the eight plots.}
\end{figure*}

Spatially resolved dI/dV spectra along a line of a topographic image
taken at 5.4 K are shown in \mbox{Fig.}
\ref{fig:spectra-along-line}. Each spectrum at a particular location is an average of 16 spectra and the distance between two consecutive locations is 2.4
nm. We clearly see that the nature of the spectra changes over nm
length scale. To see the temperature evolution of these low energy features across the T$_C$ = 15 K, we took dI/dV - V spectra at different temperatures between 5.4 K and 20 K at different locations on the surface over an area of $\sim$ 2$\times$2 $\mu$m$^2$. The spectra at 5.4 K and 19.7 K are shown in \mbox{Fig.} \ref{fig:spectra-5.4K-19.7K}. We have grouped these spectra based on the symmetry of the features observed. None of the spectra show true superconducting gap with two coherence peaks but there is an asymmetric suppression in DOS in $\pm 5$ mV bias range in some of the spectra.
In some spectra, only one peak is observed near $+ 5$ mV in the dI/dV
spectra. In some cases a sharp decrease in DOS occurs at negative
bias without any peak. Few spectra show a symmetric depression over
$\pm 5$ mV bias. In some locations the spectra show a rising DOS
with a minima near zero bias but without any low energy features,
i.e. these spectra are similar to the ones above 20 K.

\begin{figure}[h!]
\includegraphics[width=2.5in]{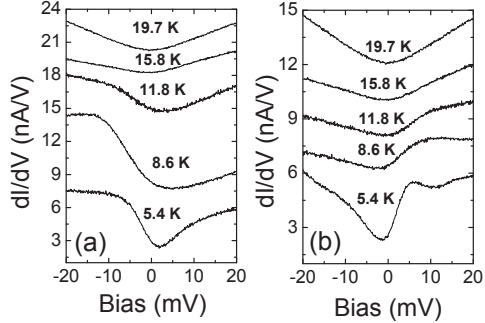}
\caption{\label{fig:spectr-all}(a) Spatially averaged dI/dV spectra
with a sharp change in DOS at -ve bias at different temperatures and
(b) Spatially averaged dI/dV spectra with a peak at + ve bias at
different temperatures. Consecutive spectra have been shifted uniformly upwards in the two plots for clarity.}
\end{figure}

Based on the observed spectra at low temperature, we divided them into three categories: spectra with a sharp change in dI/dV at -ve bias, spectra with a peak
at +ve bias, and spectra with no sharp change. We took the spatial
average of the spectra showing peak at +ve bias and of those showing
sharp change in DOS at -ve bias. We plot the temperature dependence
of these two types of spatially averaged spectra in \mbox{Fig.}
\ref{fig:spectr-all}$(a)$ and $(b)$, which demonstrate the
evolution of the DOS across T$_{C}$. Below T$_{C}$, there is
some depression in the DOS near the Fermi energy (at zero bias) and
this depression becomes more pronounced as the temperature goes
down. We would like to state that we cannot
track the same area as a function of temperature in our STM as the
relative xy-shift between tip and sample  with temperature is
significant.

Our observed tunnel spectra at low temperatures are unlike those of
typical superconductors, which show a BCS gap with two coherence
peaks in DOS. However, the observed asymmetric or symmetric
depression in DOS correlates well with the bulk superconductivity
below T$_{C}$. Above T$_{C}$, there is no peak or depression in the low bias conductance but as the temperature reduces below the superconducting
transition temperature, peak or depression in the tunnel spectra
starts to appear. The depression in DOS at low energies is most
pronounced at the lowest studied temperature. The temperature
variation of the tunnel spectra across T$_{C}$ clearly indicates
that this peak disappears above superconducting transition. If we take a spatial average of these spectra we do get a gap-like structure with symmetric depression in DOS and weak peak-like features corresponding to BCS coherence peaks. Asymmetric
spectra with gap have been seen routinely in high-T$_{C}$ cuprates
\cite{pan} with an energy gap and two coherence peaks. So the
spectra that we observed are somewhat peculiar but the low energy asymmetric features are seen only below T$_{C}$.

\section{Discussions}
Emergence of inhomogeneous DOS only below the SC onset temperature is somewhat puzzling. It is possible that some small inhomogeneities do exist above T$_C$ and
they evolve into more clearly visible features below T$_C$. These
features mainly involve an asymmetric depression in DOS near Fermi
energy in 5 meV energy range. This energy scale is consistent with
the typical superconducting gap reported in some of the pnictides
\cite{peng,jennifer}. Susceptibility data also show that a tiny fraction
($\sim$1.2$\%$) of the system is in the SC state below T$_C$.

At low temperature tunnel spectra change significantly over a few nm length scale as seen in \mbox{Fig.} \ref{fig:spectra-along-line}. A very homogeneous Ni distribution at
this doping (0.04) will give $\sim$ 1.4 nm average separation
between Ni atoms in a-b plane. So we cannot reconcile our results
with clustering of Ni atoms as that would lead to an inhomogeneity
over a larger length scale. We have also seen the same
inhomogeneous spectra in 4-5 different cleaved surfaces of these crystals,
which makes it very unlikely that Ni is segregating over length
scales larger than what is accessible to our STM. This is
further ruled out from smooth susceptibility data without any sharp
jumps. Signatures of atomic resolution and atomic steps with flat
terraces make the surface contamination a very unlikely possibility.

It is possible that the cleaved surface differs from the bulk due to reconstruction, restructuring or other disorders \cite{jennifer}. The nature of exposed surface after cleaving may depend on the cleaving temperature and a room temperature cleaving is more likely to yield a reconstructed surface. Moreover, a STM study \cite{ilija} on Pr$_x$Ca$_{1-x}$Fe$_2$As$_2$ showed three different surface morphologies, 2$\times$1, 1$\times$1 and a disordered web-like structure. Among these structures, 2$\times$1 gives a non-polar surface while 1$\times$1 gives a polar surface. In our atomic resolution images we saw only 1$\times$1 structure at some places indicating a polar surface. Due to lack of atomic resolution over large area we cannot find the surface termination at various temperatures and locations to correlate with the local spectra. Massee et. al. \cite{massee}, on room temperature cleaved BaFe$_{1.86}$Co$_{0.14}$As$_2$, showed that termination has little effect on the low-energy DOS. A hard X-ray photoemission also supports this finding \cite{jong}. Thus we find it difficult to attribute the observed inhomogeneity over a few nm length-scale, in low-energy DOS and only below T$_C$, to the spatial variation in surface termination or reconstruction.

Inhomogeneities in tunnel conductance can also arise from quasi-particle interference (QPI) as reported in a STM study on doped CaFe$_2$As$_2$ \cite{chuang}. In cuprates, spatially periodic DOS modulations have been seen over the SC gap energy scale due to QPI \cite{jennifer1}. More pronounced variations in local spectra of cuprates have been associated with oxygen inhomogeneities \cite{mcelroy}. In our spectra we see much wider spatial variations (see \mbox{Fig.} \ref{fig:spectra-along-line} and \ref{fig:spectra-5.4K-19.7K} ) than what is typically seen in QPI. We cannot rule out the QPI related inhomogeneities in present study, however we find it difficult to completely attribute these inhomogeneities to QPI.

In a typical pnictide phase diagram similar to cuprate superconductors, there is the famous superconducting dome and a phase boundary extending to much higher
temperatures that separates the SDW phase from the paramagnetic
phase. Thus the phase boundaries touch the T = 0 axis at three
points and it is not fully clear if all the
three points are QCPs. The temperature dependent resistivity,
penetration depth and spin-lattice relaxation measurements have
strongly suggested that the point inside the dome is a QCP in pnictides
\cite{zhou,shibauchi}. In cuprates, field doping near the first point (on underdoped side) has also revealed a QCP due to crossover between SC phase and an
insulator phase \cite{bollinger}. Larger inhomogeneities have been
observed in underdoped cuprates than the overdoped ones
\cite{gomes}. Superconductor to insulator transition has also been
reported in heavily disordered conventional superconductors although
it is not clear if this disorder also acts as a dopant and directly
affects the SC order \cite{goldman}.

The composition of our crystals is close to the superconducting dome boundary on the
underdoped side. Proximity to this boundary makes the non-SC phases
easily accessible and presence of disorder will further help in
nucleating such phases. We believe that in proximity to this
crossover point, which may be a QCP, the system will be extremely
susceptible to disorder and eventually the disorder might influence
this crossover more than doping. Our results could be suggestive of the
presence of a QCP at a point where the phase boundary touches the temperature axis in the underdoped side of pnictides.

\section{Conclusions}
In conclusion, our temperature dependent STM/S
investigation of underdoped CaFe$_{1.96}$Ni$_{0.04}$As$_{2}$ single
crystals show a highly inhomogeneous electronic phase below the SC transition temperature. At high temperatures we see a SDW gap consistent with the resistivity measurements. Magnetic measurement show that only a tiny fraction ($\sim$ 1.2 $\%$) of this compound superconducts below 15 K. Our STM measurements show atomically flat terraces with signatures of atomic resolutions at lowest temperature. Above T$_C$,
the tunnel spectra show homogeneous local DOS with a SDW gap. But, below T$_C$, low energy scale spectra at some locations show an asymmetric or a symmetric dip in DOS at $\pm 5$ meV energy, while, at other locations the spectra are similar to the ones at higher temperatures with SDW gap. This asymmetric or symmetric depression correlates well with the superconducting phase and the inhomogeneity disappears as the temperature goes above the superconducting transition temperature. Observed inhomogeneity over such a few nm length-scale and smooth susceptibility data also exclude the possibility of segregation of Ni. From this STM/S study we believe that the inhomogeneities in this underdoped compound below T$_C$ are intrinsic due to the proximity to the non-SC phase.

\section*{Acknowledgements}
We thank Sourabh Barua for his help in the resistivity measurement.
We also like to acknowledge Amit Dutta for helpful discussions.
Anirban acknowledges financial support from the CSIR of the
Government of India. A.K.G. acknowledges a research grant from the
CSIR of the Government of India.

\end{document}